\documentclass[aps,prl,twocolumn,showpacs]{revtex4}
\usepackage{epsfig}
\begin{document}
\title{Partonic Effects on Pion Interferometry at the Relativistic Heavy Ion Collider}
\author{Zi-wei Lin, C.M. Ko, and Subrata Pal}
\affiliation{Cyclotron Institute and Physics Department,
Texas A\&M University, College Station, Texas 77843-3366}
\begin{abstract}
Using a multiphase transport (AMPT) model that includes both initial 
partonic and final hadronic interactions, we study the pion interferometry 
at the Relativistic Heavy Ion Collider. We find that the two-pion 
correlation function is sensitive to the magnitude of the parton scattering 
cross section, which controls the parton density at which the transition 
from the partonic to hadronic matter occurs. Also, the emission source of pions
is non-Gaussian, leading to source radii that 
can be more than twice larger than the radius parameters extracted from 
a Gaussian fit to the correlation function.  
\end{abstract}
\maketitle

Particle interferometry based on the Hanbury-Brown Twiss (HBT) effect
has long been used to measure the size of an emission source \cite{hbt}. 
In heavy ion collisions, it has been suggested that the HBT study can 
provide information not only on the spatial extent of the emission source 
but also on its expansion velocity and emission duration 
\cite{Pratt:su,Bertsch:1988db,Pratt:zq,Rischke:1996em}. In particular, 
the long emission time as a result of the phase transition from the 
quark-gluon plasma to hadronic matter in relativistic heavy ion collisions 
is expected to lead to an emission source which has a much larger radius 
in the direction of the total transverse momentum of detected two 
particles ($R_{\rm out}$) than that perpendicular to both this direction and 
the beam direction ($R_{\rm side}$) 
\cite{Rischke:1996em,Soff:2000eh,Soff:2001hc}. 
Since the quark-gluon plasma is expected to be formed in heavy
ion collisions at the Relativistic Heavy Ion Collider (RHIC), 
it is thus surprising to find that the extracted 
ratio $R_{\rm out}/R_{\rm side}$ from a Gaussian fit to the measured
two-pion correlation function in Au+Au collisions 
is close to one \cite{Adler:2001zd,Johnson:2001zi,Adcox:2002uc}. 
Also, the extracted radius parameters are small compared to  
theoretical predictions based on the hydrodynamical model \cite{Soff:2000eh}.  
These experimental results, especially the small value of 
$R_{\rm out}/R_{\rm side}$, have been attributed to strong space-time 
and momentum correlations in the emission source \cite{Tomasik:1998qt}.  

In this Letter, a multiphase transport model (AMPT) 
\cite{Zhang:2000bd,Lin:2001cx,Lin:2001zk}, that includes
both initial partonic and final hadronic interactions, is used to study 
the pion interferometry in central Au+Au collisions at RHIC at 
$\sqrt s=130A$ GeV. The AMPT model is a hybrid model that 
uses the minijet partons from hard processes and the strings 
from soft processes in the HIJING model \cite{Wang:1991ht}
as the initial conditions for modeling the collision dynamics. 
The time evolution of partons is then modeled by the ZPC \cite{Zhang:1997ej}
parton cascade model. At present, this model includes only parton-parton 
elastic scatterings with an in-medium cross section given by:
\begin{eqnarray}
\frac{d\sigma_p}{d\hat t}=\frac{9\pi\alpha_s^2}{2} 
\left (1+{{\mu^2} \over \hat s} \right ) \frac{1}{(\hat t-\mu^2)^2}, 
\end{eqnarray}
where the strong coupling constant $\alpha_s$ is taken to be 0.47, and
$\hat s$ and $\hat t$ are the Mandelstam variables.  The effective screening 
mass $\mu$ depends on the temperature and density of the partonic matter
but is taken as a parameter for fixing the magnitude and
angular distribution of the parton scattering cross section.
After these minijet partons stop interacting, they are combined
with their parent strings to fragment to hadrons using
the Lund string fragmentation model as implemented in the PYTHIA routine 
\cite{Sjostrand:1994yb}. The final-state hadronic scatterings are 
then modeled by the ART model \cite{Li:1995pr}. 

The default AMPT model has been quite successful 
in describing the measured rapidity distributions of charge particles, 
particle to antiparticle ratios, and the spectra of low transverse momentum 
pions and kaons \cite{Lin:2001cx} in heavy ion collisions at 
the Super Proton Synchrotron (SPS) and RHIC. 
Since the initial energy density in Au+Au collisions at RHIC 
is expected to be much larger than the critical energy density 
at which the hadronic matter to quark-gluon plasma transition would
occur \cite{Kharzeev:2001ph,Zhang:2000nc}, the AMPT model has been 
extended to convert the initial excited strings into partons 
\cite{Lin:2001zk}. In this string melting scenario, hadrons, that 
would have been produced from string fragmentation, are converted instead
to valence quarks and/or antiquarks. Interactions among these 
partons are again described by the ZPC parton cascade model. 
The transition of the partonic matter to the hadronic matter is,
however, achieved using a simple coalescence model, which combines
two nearest partons into mesons and three nearest partons into 
baryons or anti-baryons. Using parton scattering cross sections of 
6-10 mb, the extended AMPT model is able to reproduce both the centrality 
and transverse momentum (below 2 GeV$/c$) dependence of the elliptic flow 
measured in Au+Au collisions at $\sqrt s=130A$ GeV at RHIC 
\cite{Ackermann:2000tr}.

From the AMPT model, the source of emitted particles is obtained from
their space-time coordinate $x$ and momentum ${\bf p}$ at freezeout, 
i.e., at their last interactions. Denoting the single-particle 
emission function for pions by $S(x,{\bf p})$, the HBT correlation function 
for two identical pions 
in the absence of final-state interactions, such as the Coulomb 
interaction, is then given by \cite{Pratt:su,Wiedemann:1999qn}:
\begin{eqnarray}\label{emission}
C_2(\!{\bf Q},\!{\bf K}\!)\!\!=\!\!1\!\!+\!\!
\frac{\int \!\!d^4\!x_1d^4\!x_2 S(x_1,\!{\bf K}\!)
S(x_2,\!{\bf K}\!) \cos \!\left [Q \!\! \cdot \! (x_1\!\!-\!\!x_2) \right ]}
{\int d^4x_1 S(x_1,{\bf p_1}) \int d^4x_2 S(x_2,{\bf p_2})},
\end{eqnarray}
where ${\bf p_1}$ and ${\bf p_2}$ represent the momenta of the two pions, 
respectively, ${\bf K}=({\bf p_1}+{\bf p_2})/2$, 
and $Q=({\bf p_1}-{\bf p_2}, E_1-E_2)$. 
Expecting that the emission function is sufficiently smooth 
in the momentum space, one can evaluate the correlation function by using
${\bf p_1}$ and ${\bf p_2}$ for ${\bf K}$ in the numerator
of Eq.(\ref{emission}). Furthermore, the three-dimensional correlation 
function in ${\bf Q}$ is usually shown as a function of the invariant 
relative momentum ($Q_{\rm inv}=\sqrt{-Q^2}$) or as a function of the 
projection of ${\bf Q}$ in the ``out-side-long'' 
(${\it osl}$) system \cite{Bertsch:1988db,Pratt:zq}, defined by the 
beam direction ($Q_{\rm long}$), the direction along the total momentum 
of the two particles in the transverse plane ($Q_{\rm out}$), and the 
direction orthogonal to the above two directions ($Q_{\rm side}$).

\begin{figure}[h]
\centerline{\epsfig{file=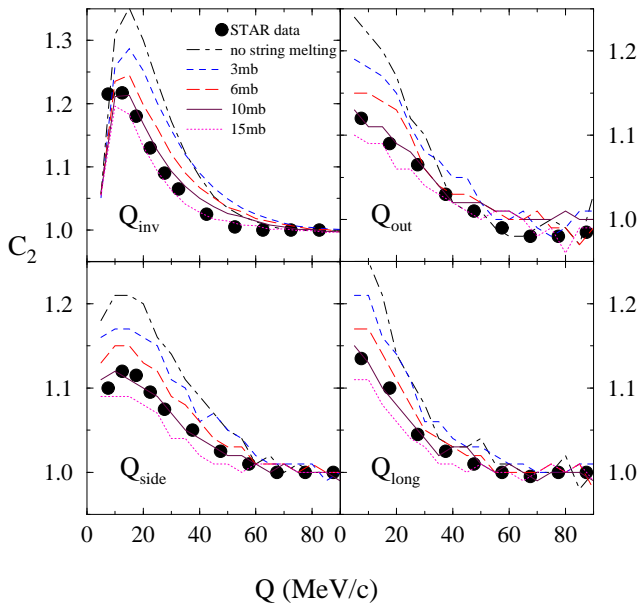,width=3.3in,angle=0}}
\vspace{0.5cm}
\caption{Correlation functions for midrapidity charged pions with 
$125<p_{\rm T}<225$ MeV$/c$. Theoretical results with Coulomb corrections 
are shown for the default AMPT model 
and for the extended AMPT model with string melting and various values 
for $\sigma_p$. Coulomb-uncorrected correlation functions from 
the STAR collaboration \protect\cite{Adler:2001zd} are shown 
by filled circles.} 
\label{cpim}
\end{figure}

Using the emission function obtained from the AMPT model for central 
($b=0$ fm) Au+Au collisions at $\sqrt{s}=130A$ GeV, we have evaluated 
the correlation function $C_2({\bf Q},{\bf K})$ in the longitudinally 
comoving system frame using the program Correlation After Burner 
\cite{pratt:uf}. In Fig. \ref{cpim}, we show the one-dimensional projections
of the calculated correlation function including final-state Coulomb 
interactions for midrapidity ($-0.5<y<0.5$) charged pions with transverse 
momentum $125<p_{\rm T}<225$ MeV$/c$. Also shown are the measured $\pi^-$ 
correlation functions from central collisions by the STAR collaboration 
without removing the effect due to the Coulomb interaction 
\cite{Adler:2001zd}. In evaluating the one-dimensional projections  
of the correlation function onto one of the $Q_{\rm out}, Q_{\rm side}$ 
and $Q_{\rm long}$ axes, we have integrated the other two ${\bf Q}$ 
components over the range $0-35$ MeV/$c$. The dash-dotted curves in 
Fig. \ref{cpim} are results from the default AMPT model (no string melting) 
with a parton scattering cross section of $\sigma_p=3$ mb, while other curves 
are those from the extended AMPT model with string 
melting but different values for $\sigma_p$. It is seen that with string 
melting both the width of the $Q_{\rm inv}$ correlation function and its 
height decrease with increasing $\sigma_p$. The decreasing width 
can be more clearly seen from the calculated correlation functions 
without including final-state Coulomb interactions \cite{long}, as 
the value of the correlation function in this case is exactly two at 
$Q_{\rm inv}=0$. Since the integration of the other two ${\bf Q}$ 
components over a fixed range ($0-35$ MeV/$c$) gives a smaller value when 
the correlation function becomes narrower in ${\bf Q}$, the height of 
the one-dimensional projection of the correlation function decreases 
with increasing $\sigma_p$. To reproduce the measured 
one-dimensional correlation functions by the STAR collaboration, 
we find that a parton scattering cross section of about 10 mb  
is required in the extended AMPT model with string melting.
Since in our model partons are converted to hadrons after they 
make their last scattering, the parton scattering cross section
thus controls the density at which partons are converted to hadrons.
A larger $\sigma_p$ leads to a lower density and larger volume 
for the parton to hadron phase transition.  We expect that 
the two-pion correlation data could also be reproduced by
using a smaller parton scattering cross section but delaying the
hadronization until the critical density is similar to the one
given by the parton freezeout condition using a scattering 
cross section of 10 mb in the parton cascade model \cite{long}.

The size of the emission source can be determined from the emission
function via the curvature of the correlation function at ${\bf Q}=0$:
\begin{eqnarray}
&&R_{ij}(K)^2=\left . -{1 \over 2}
\frac{\partial^2 C_2({\bf Q}, {\bf K})}{\partial Q_i \partial Q_j} 
\right |_{{\bf Q}=0} \nonumber \\
&&=D_{x_i,x_j}(\!K\!)\!\!-\!\!D_{x_i,\beta_j t}(\!K\!)\!\!
-\!\!D_{\beta_i t,x_j}(\!K\!)
\!\!+\!\!D_{\beta_i t,\beta_j t}(\!K\!),
\label{source}
\end{eqnarray}
where $x_i(i=1-3)$ denotes the projections of the particle position at 
freezeout in the {\em osl} system, i.e., $x_{\rm out}$, $ x_{\rm side}$ 
and $x_{\rm long}$, and ${\bf \beta}={\bf K}/K_0$ with $K_0$ being 
the average energy of the two particles.  In the second line of 
Eq.(\ref{source}), we have the variance 
$D_{x,y}=\langle x\cdot y \rangle-\langle x\rangle \langle y \rangle$, 
with $\langle x\rangle$ denoting the average value of $x$.  

Since only the correlation function is measured in experiments, the 
size of emission source is usually estimated by fitting the measured
correlation function $C_2({\bf Q},{\bf K})$, that has been corrected 
for effects due to final-state Coulomb interactions, with a four-parameter 
Gaussian function:
\begin{eqnarray}
C_2({\bf Q},{\bf K})=1+ \lambda \exp 
\left ( -\sum_{i=1}^3 R^2_{ii}(K) Q_i^2 \right ). 
\label{hbt}
\end{eqnarray}
If the emission source is Gaussian in space-time, then for central heavy 
ion collisions considered here the radius parameters obtained from 
the above Gaussian fit to the correlation function would be the same 
as those determined directly from the emission function of the source 
via Eq.(\ref{source}). However, because of space-time correlations in 
the emission function, such as those induced by the collective flow, 
the radius parameters in general do not give the true source size. 

\begin{figure}[h]
\centerline{\epsfig{file=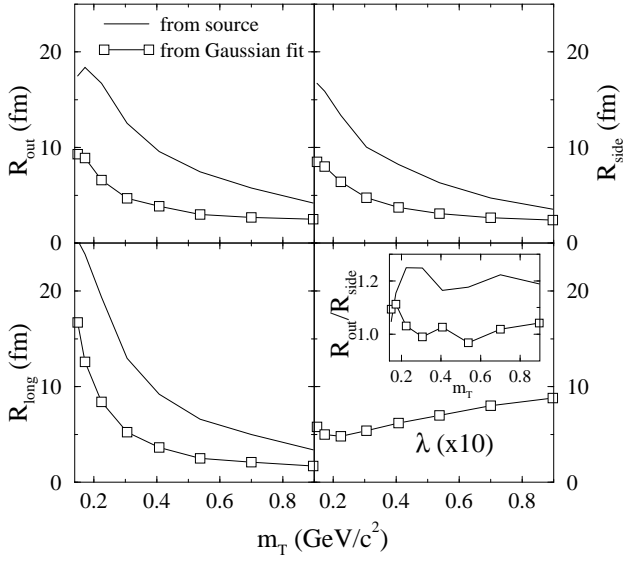,width=3.3in,angle=0}}
\vspace{0.5cm}
\caption{Source radii from the emission function (solid curves) as well
as the fitted radius and $\lambda$ parameters from the Gaussian 
fit to the correlation function (curves with squares)
for midrapidity pions as functions of pion transverse mass $m_{\rm T}$. 
The inset shows the corresponding ratio $R_{\rm out}/R_{\rm side}$.}
\label{rkt}
\end{figure}

In Fig. \ref{rkt}, we show by solid curves the transverse mass ($m_{\rm T}$) 
dependence of the source radii $R_{\rm out}$ (upper-left panel), 
$R_{\rm side}$ (upper-right panel), and $R_{\rm long}$ (lower-left panel) 
determined from the emission function for midrapidity charged pions 
given by the AMPT model with string melting and parton cross section 
of $\sigma_p=10$ mb for central Au+Au collisions at $\sqrt s=130A$ GeV. 
The radius parameters determined from the Gaussian fit to the 
three-dimensional correlation function in ${\bf Q}$ 
without including final-state Coulomb interactions
are shown by curves with squares. For the $m_{\rm T}$ values shown here, 
these radius parameters are about a factor of 2 to 3 smaller than the 
source radii obtained directly from the emission function. The emission 
source from the AMPT model thus deviates appreciably from a Gaussian one.
The radius parameters from the Gaussian fit to the correlation function
have similar values as the experimental ones from a Gaussian fit to
the measured correlation function after correcting for the final-state
Coulomb interactions \cite{Adler:2001zd}. The parameter $\lambda$ 
(scaled up by a factor of 10) from the Gaussian fit to the correlation 
function is shown in the lower-right panel by the curve with squares. 
It has a value of about 0.5 at low $m_{\rm T}$ but increases to about 1
at large $m_{\rm T}$.  Both the source radii and fitted radius parameters 
decrease with increasing $m_{\rm T}$. On the other hand, these radii 
increase with increasing parton scattering cross section as a result of 
both a larger source size and stronger collective expansion \cite{long}. 

Since Eq.(\ref{source}) gives 
\begin{eqnarray}
R_{\rm out}^2&=&D_{x_{\rm out},x_{\rm out}}-2~D_{x_{\rm out},\beta_\perp t}
+D_{\beta_\perp t,\beta_\perp t},
\label{rout}
\end{eqnarray}
and $R_{\rm side}^2=D_{x_{\rm side},x_{\rm side}}$, the ratio 
$R_{\rm out}/R_{\rm side}$ contains information about 
the duration of emission and has been studied extensively 
\cite{Soff:2000eh,Adler:2001zd}. This connection between $R_{\rm out}$ 
and the emission duration becomes clearer if we neglect the 
$x_{\rm out}-t$ correlation term $D_{x_{\rm out},\beta_\perp t}$ 
in Eq.(\ref{rout}). In the inset of the lower-right panel of 
Fig. \ref{rkt}, we show the ratio $R_{\rm out}/R_{\rm side}$ for 
midrapidity charged pions. It is seen that the ratio 
$R_{\rm out}/R_{\rm side}$ obtained from the emission function 
(solid curve) has a value between 1.0 and 1.3 as in predictions based 
on the hydrodynamical model with freezeout treated via the hadronic 
transport model \cite{Soff:2000eh}. However, with the radius parameters 
extracted from the Gaussian fit to the correlation function, the ratio
$R_{\rm out}/R_{\rm side}$ becomes much closer to 1, similar to 
the experimental values extracted from the measured correlation
function \cite{Adler:2001zd}. 

\begin{figure}[h]
\centerline{\epsfig{file=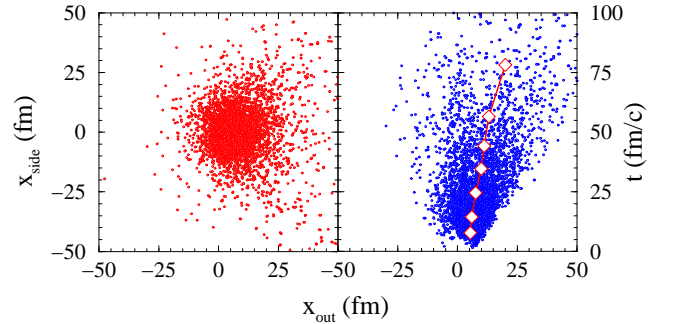,width=3.3in,angle=0}}
\vspace{0.5cm}
\caption{$x_{\rm out}-x_{\rm side}$ (left panel) and $x_{\rm out}-t$ 
(right panel) distributions at freezeout for midrapidity charged pions 
with $125<p_{\rm T}<225$ MeV$/c$.
The curve with open diamonds represents $\langle x_{\rm out} \rangle$ 
as a function of $t$.}
\label{outsidet}
\end{figure}

To investigate the reason for the large difference between 
the source radii obtained directly from the emission function
and from the Gaussian fit to the correlation function, we show
in Fig. \ref{outsidet} the $x_{\rm out}-x_{\rm side}$ distribution 
(left panel) and the $x_{\rm out}-t$ distribution (right panel) 
at freezeout for midrapidity pions with $125<p_{\rm T}<225$ MeV$/c$ 
from the AMPT model with string melting and $\sigma_p=10$ mb.
It is seen that the emission source is shifted in the direction of
the pion transverse momentum, i.e, $\langle x_{\rm out} \rangle >0$. 
This positive shift in $x_{\rm out}$ results from the collective 
expansion of the emission source \cite{Lin:2001zk}.

If the width in $x_{\rm out}$ were much smaller than the average
value $\langle x_{\rm out} \rangle$, the emission source would 
have a shell-like shape, similar to the emission source with strong opacity 
in the hydrodynamical model \cite{Tomasik:1998qt}. 
The emission source also shows a large halo around a central core.  
The halo consists not only of pions from decays of 
long-lived resonances such as the $\omega$ but also of thermal pions. 
In calculating the correlation function, we have included pions from 
the decay of $\eta$ resonances as in experiments. 
Their effects on the radius parameters obtained from the Gaussian
fit to the correlation function are thus included. On the other hand, 
we have excluded these pions in evaluating the source radii from the 
emission function due to the long lifetime of $\eta$.
Since long-lived resonances mainly affect the correlation 
function at small relative momenta \cite{Csorgo:1994in}, 
they are important in determining the $\lambda$ parameter. 

As to the $x_{\rm out}-t$ distribution of the emission function shown in the 
right panel of Fig. \ref{outsidet}, it has a strong positive $x_{\rm out}-t$ 
correlation as clearly seen from the solid curve with open diamonds,
which shows that the average value $\langle x_{\rm out} \rangle$ increases
with the freezeout time $t$. The $x_{\rm out}-t$ correlation leads to a large 
positive value for the $x_{\rm out}-t$ correlation term $D_{x_{\rm out}, 
\beta_\perp t}$ in Eq.(\ref{rout}).  For pions included in generating 
Fig. \ref{outsidet}, the value of $D_{x_{\rm out}, \beta_\perp t}$ is 
168 fm$^2$ and is appreciable compared to 185 and 431 fm$^2$, 
respectively, for the first and last terms in Eq.(\ref{rout}).  
This makes it difficult to extract information about the duration of 
emission from the ratio $R_{\rm out}/R_{\rm side}$. Similar results 
have been seen previously in studies based on the RQMD model 
at SPS \cite{Sullivan:wb,Fields:sj}. We note that the 
imaging method \cite{imaging}, developed for extracting the emission 
function of a source from the correlation function, will be very 
useful for verifying the non-Gaussian features of the emission source 
in high energy heavy ion collisions.

In summary, we have studied the pion interferometry in relativistic 
heavy ion collisions at RHIC. Using a multiphase transport model 
that includes both initial partonic and final hadronic
interactions, we find that the two-pion correlation function is 
sensitive to the parton scattering cross section, which controls the 
density at which the parton-to-hadron transition occurs in the AMPT 
transport model. To reproduce the measured correlation function in central 
Au+Au collisions at $\sqrt s=130A$ GeV requires both the melting of initial
strings to partons and a large parton scattering cross section, 
i.e., a low parton density for the transition from the partonic to 
the hadronic matter. We further find that the emission source 
is non-Gaussian in space and time. It not only shifts significantly to 
the direction along the pion transverse momentum but also has a strong 
correlation between this displacement and the freezeout time. 
Consequently, the source radii extracted directly from the emission 
function are about a factor of 2 to 3 larger than the radius parameters 
extracted from a Gaussian fit to the three-dimensional correlation function. 
Furthermore, the ratio $R_{\rm out}/R_{\rm side}$ obtained from the emission 
function is larger than that extracted from a Gaussian fit to the 
correlation function, which is found to be close to one. Although 
the correlation function requires only the space-time information 
of pions at freezeout, it is shown to be  sensitive to the partonic
dynamics during the early stage of heavy ion collisions. The study of pion 
interferometry thus helps to confirm the formation of the partonic
matter at RHIC and to study its properties.

\begin{acknowledgements}
We appreciate useful discussions with L.-W. Chen, P. Danielewicz,
M. Gyulassy, U. Heinz, M. Lisa, M. Murray, P. Philip, T. Cs\"org\"o, S. Pratt, 
N. Xu, and Q.H. Zhang. This paper is based on work supported by 
the U.S. National Science Foundation under Grant Nos. PHY-9870038 
and PHY-0098805, the Welch Foundation under Grant No. A-1358, and 
the Texas Advanced Research Program under Grant No. FY99-010366-0081.
\end{acknowledgements}

\end{document}